\documentclass[twoside,a4paper,11pt]{sea10}

\usepackage{graphicx}
\usepackage{hyperref}
\usepackage{movie15}
\usepackage[numbers,square,sort&compress]{natbib}

\def\specchar#1{{\sc #1}}

\def\degree{\hbox{$^\circ$}}
\def\arcsec{\hbox{$^{\prime\prime}$}}
\def\FeI{\mbox{Fe\,\specchar{i}}}
\def\FeII{\mbox{Fe\,\specchar{ii}}}

\def\HeI{\mbox{He\,\specchar{i}}}
\def\CaII{\mbox{Ca\,\specchar{ii}}}

\def\Halpha{\mbox{H$\alpha$}}

\def\kms{\hbox{km$\;$s$^{-1}$}}

\DeclareMathAlphabet{\mathsc}{OT1}{cmbr}{n}{sc}

\begin{document}
\pagenumbering{arabic}
\pagestyle{myheadings}
\thispagestyle{empty}
{\flushleft\includegraphics[width=\textwidth,bb=58 650 590 680]{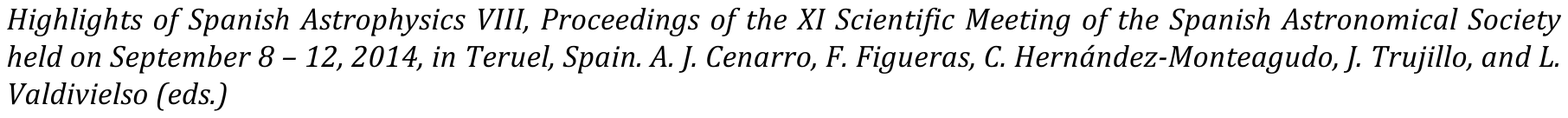}}
\vspace*{0.2cm}
\begin{flushleft}
{\bf {\LARGE
Beyond MHD: modeling and observation of partially ionized solar plasma processes
}\\
\vspace*{1cm}
E. Khomenko$^{1, 2}$
}\\
\vspace*{0.5cm}
$^{1}$
Instituto de Astrof\'{\i}sica de Canarias, 38205 La Laguna, Tenerife, Spain\\
$^{2}$
Dpto de Astrof\'{\i}sica, Universidad de La Laguna, 38205, La Laguna, Tenerife, Spain
\end{flushleft}
%
\markboth{}{ 
%
E. Khomenko
%
}
\thispagestyle{empty}
\vspace*{0.4cm}
\begin{minipage}[l]{0.09\textwidth}
\ 
\end{minipage}
\begin{minipage}[r]{0.9\textwidth}
\vspace{1cm}
\section*{Abstract}{\small
The temperature and density conditions in the magnetized photosphere and chromosphere of the Sun lead to a very small degree of atomic ionization. In addition, at particular height, the magnetic field may be strong enough to give rise to a cyclotron frequency larger than the collisional frequency for some species, while for others the opposite may happen. These circumstances influence the collective behavior of the particles and some of the hypotheses of magnetohydrodynamics may be relaxed, giving rise to non-ideal MHD effects. In this paper we discuss our recent developments in modeling non-ideal plasma effects derived from the presence of a large amount of neutrals in the solar photosphere and the chromosphere, as well as observational consequences of these effects. 
\normalsize}
\end{minipage}
\section{Introduction \label{intro}}

The plasma in the photosphere and chromosphere of the Sun is often treated in the approximation of magnetohydrodynamics (MHD). MHD modeling has been very successful both in constructing idealized models of basic processes, and in providing realistic models of complex solar phenomena such as magneto-convection, formation of magnetic structures, flux emergence, wave propagation, reconnection, etc. \citep{Asplund+etal2000, Khomenko+Collados2006, Cheung+etal2007, Moreno-Insertis+etal2008}. While in ideal MHD the plasma is supposed to be completely ionized, the solar plasma is only weakly ionized, and the ionization degree can drop as  low as 10$^{-4}$ in the upper photosphere. Together with the decrease of the collisional coupling with height, this may lead  to the break of magnetohydrodynamic assumptions (depending on temporal and spatial scales of the phenomena under consideration) and lead to a series of non-ideal plasma effects. Since the degree of magnetization (the ratio of cyclotron to collision frequency of the different species) depends on the magnetic field, it is expected that non-ideal effects may play role in strongly magnetized photospheric regions (sunspots, flux tubes) and all over the chromosphere. 

In order to take into account the non-ideal effects derived from the presence of neutrals, a fluid-like description has often been used in the photosphere and the chromosphere. This is possible because the collisions are still strong enough to partially couple different plasma components. Many of  single-fluid and multi-fluid models include only hydrogen or hydrogen-helium plasma, and a simplified treatment of radiative energy exchange (or even without taking radiative transfer aspects into account), which greatly simplifies the complexity of the problem. Such an approach has been frequently used in studies of different types of waves in the solar atmosphere \cite{Khodachenko2004, Forteza2007, Vranjes2008, Soler2009, Zaqarashvili+etal2012} and studies of instabilities \citep{Soler+etal2012, Diaz+etal2012, Diaz+etal2013}. Idealized models of reconnection have been constructed in single-fluid and two-fluid approaches  \citep{Zweibel1989,  Leake+etal2012, Sakai2006}. More complex models include those of magnetic flux emergence \citep{Leake+Arber2006, Arber2007} and of chromospheric thermal structure \citep{Krasnoselskikh2010, Khomenko+Collados2012, MartinezSykora+etal2012}. The presence of neutrals, via collisions, provides a source of additional dissipation and energy supply. A weakly ionized plasma is able to diffuse through the magnetic field, partially breaking the frozen-in condition. The influence of neutrals results in creation or removal of cut-off frequencies for waves and instabilities, additional heating by dissipation of waves and currents, anisotropic dissipation of currents perpendicular to magnetic field facilitating the generation of potential structures, and other phenomena. Therefore, the influence of neutrals on the dynamics and energetics of solar plasma is proved to be important.

With a constant increase of resolution capabilities of solar observations, both from space and from the ground, it is  essential that new realistic models are built including non-ideal effects derived from the presence of neutrals. In this paper, we describe the efforts undertaken by our group in the development of an analytical theory and a numerical code for such modeling. We attempt to provide a self-consistent description of a multi-component multi-species solar plasma and its interaction with the radiation field. Direct observational detection of ion-neutral effects has also been attempted and its first results are described.

\section{Equations for partially ionized solar plasma}

We consider a plasma composed of a mixture of atoms of different atomic species. These atoms can be excited to excitation levels and/or to different ionization stages. Without loss of generality we limit the ionization stages to only two, as the number of multiply ionized ions is small in the regions of interest of the solar atmosphere.  The interaction with radiation is included via excitation/deexcitation and ionization/recombination processes and photons are treated as another type of particles interacting with the rest of the mixture. The detailed derivation of the macroscopic equations of continuity, momentum and energy for each ionization-excitation level of species composing solar plasma components is given in \cite{Khomenko+etal2014b}. After summing up the motion equations over excitation states, a two-fluid system of equations is derived that describes the behavior of neutral and ionized species. The electron and ion contributions are added up together to get a single equation for the charged component.  This way, the following system of equations is obtained:

\begin{equation}  \label{eq:rhon}
\frac{\partial \rho_n}{\partial t} + \vec{\nabla} (\rho_n\vec{u}_n) = S_n
\end{equation}
\begin{equation} \label{eq:rhoi}
\frac{\partial \rho_c}{\partial t} + \vec{\nabla} (\rho_c\vec{u}_c) = -S_n
\end{equation}
\begin{equation} \label{eq:rhovn}
\frac{\partial (\rho_n\vec{u_n})}{\partial t} + \vec{\nabla}(\rho_n\vec{u_n} \otimes \vec{u_n})= \rho_n\vec{g} -\vec{\nabla}{\bf\hat{p}}_n+\vec{R}_n
\end{equation}
\begin{equation} \label{eq:rhovi}
\frac{\partial (\rho_c\vec{u_c})}{\partial t} + \vec{\nabla}(\rho_c\vec{u}_c\otimes\vec{u}_c)=[\vec{J}\times\vec{B}] + \rho_c\vec{g}-\vec{\nabla} {\bf\hat{p}}_{ie} -\vec{R}_n
\end{equation}
\begin{eqnarray} \label{eq:en}
\frac{\partial }{\partial t}\left(e_n+\frac{1}{2}\rho_n u_n^2\right) + \vec{\nabla}\left(\vec{u}_n (e_n+\frac{1}{2}\rho_n u_n^2) + {\bf\hat{p}}_n\vec{u}_n +\vec{q}_n^{\,\, \prime} \right) =\rho_n\vec{u}_n\vec{g} + M_n - \vec{\nabla}\vec{F}_R^n  
\end{eqnarray}
\begin{eqnarray} \label{eq:ei}
\frac{\partial }{\partial t}\left(e_{ei}+\frac{1}{2}\rho_c u_c^2\right) + \vec{\nabla}\left(\vec{u}_c(e_{ei}+\frac{1}{2}\rho_c u_c^2 ) +{\bf\hat{p}}_{ie}\vec{u}_c + \vec{q}_{ie}^{\,\, \prime} \right)  =  \rho_c\vec{u}_c\vec{g}  +\vec{J}\vec{E}^* -M_n -\vec{\nabla}\vec{F}_R^c
\end{eqnarray}

\noindent In the above equations $\rho_n =\sum_\alpha n_{\alpha n}m_{\alpha n}$ and $\rho_c=\sum_\alpha n_{\alpha i}m_{\alpha i} + n_em_e$ are the total mass density of neutrals and charges; $\vec{u}_n $ and are $\vec{u}_c$ their center of mass velocities; $\vec{J}$ is the total current density; $e_n=3p_n/2 + \sum_\alpha \chi_{\alpha n} $ and  $e_{ei}=3p_{ei}/2 + \sum_\alpha \chi_{\alpha i} $ are the internal energies of neutrals and ions, where $\chi_{\alpha n}$ and $\chi_{\alpha i}$ are the ionization energies of species $\alpha$. The summation in $\alpha$ goes over $N$ chemical components.
The definition of neutral, ${\bf\hat{p}}_n$, and  electron-ion, ${\bf\hat{p}}_{ie}$,  pressure tensors and heat flux vectors, $\vec{q}_n^{\,\, \prime}$ and $\vec{q}_{ie}^{\,\, \prime}$, is done using a common system of reference for neutral and charged velocities, see \cite{Khomenko+etal2014b}.  The heat flow vectors are then corrected for the potential ionization-recombination energy flux, e.g.  $\vec{q}^{\,\, \prime}_n  =  \vec{q}_n + \sum_\alpha \chi_{\alpha n} \vec{w}_{\alpha n}$. In this expression, the drift velocities of a neutral/ion of species $\alpha$ (e.g., $\vec{w}_{\alpha n}$) are taken with respect to the central of mass velocity of all neutrals/ions, e.g. $\vec{w}_{\alpha n}=\vec{u}_{\alpha n}-\vec{u}_n$. 
For the closure, one needs to provide the  for collisional mass, momentum and energy exchange, $S_n$, $\vec{R}_n$ and $M_n$. The mass source term $S_n$ is due to imbalance between the ionization/recombination processes and depends on the number densities and on the collisional rate coefficients. The term $M_n$ accounts for the energy exchange of neutrals through the elastic collisions with charges. Due to their complexity in a general case, we do not provide specific expressions for these terms. The collisional momentum exchange can be simplified to give \cite{Khomenko+etal2014b}:
\begin{eqnarray}
\vec{R}_n \approx &-&\rho_e(\vec{u}_n - \vec{u}_e) \sum_{\beta=1}^N \nu_{en_\beta} - \rho_i(\vec{u}_n - \vec{u}_c) \sum_{\alpha=1}^N\sum_{\beta=1}^N\nu_{i_\alpha n_\beta}
\end{eqnarray}
where $\nu_{en_\beta}$ are the collisional frequencies between electrons and neutrals of species $\beta$ and $\nu_{i_\alpha n_\beta}$ are those between ions and neutrals of different species. 

The energy exchange by photons in Eqs. \ref{eq:en} and \ref{eq:ei} is split into the contributions due to neutrals and charges to the radiative energy flux as follows:
\begin{equation}
\vec{F}_R = \vec{F}_R^n + \vec{F}_R^c
\end{equation}
where 
\begin{eqnarray} 
\vec{\nabla}\vec{F}_R^n =  \int_0^{\infty} \oint (j_\nu^n - k_\nu^n I_\nu) d\Omega d\nu; \,\,\,\,\,\,\,\,
\vec{\nabla}\vec{F}_R^c =  \int_0^{\infty} \oint (j_\nu^c - k_\nu^c I_\nu) d\Omega d\nu 
\end{eqnarray}
In these definitions we have separated the contributions in the absorption $k_\nu$ and emission $j_\nu$ coefficients related to neutrals and ions. The intensities  $I_\nu$ are obtained from the complete radiative transfer equation, with total coefficients $j_\nu$ and $k_\nu$.
\begin{equation} \label{eq:transfer}
\frac{d\,I_\nu}{d\,s} = j_\nu - k_\nu I_\nu
\end{equation}

\noindent To close the system (\ref{eq:rhon}--\ref{eq:ei}), we use the generalized Ohm's law providing the electric field, $\vec{E}$:
\begin{eqnarray}
\vec{E}^*=[\vec{E} + \vec{u}_c\times{B}] = \frac{\rho_e}{(en_e)^2}\left(\sum_\alpha \nu_{ei_\alpha} + \sum_\beta \nu_{en_\beta} \right)\vec{J} + \frac{1}{en_e}[\vec{J}\times \vec{B}] -\frac{\vec{\nabla}{\bf\hat{p}}_e}{en_e}  \\ \nonumber
- \frac{\rho_e}{en_e}(\vec{u}_c - \vec{u}_n)\left( \sum_\beta\nu_{en_\beta} - \sum_\alpha\sum_\beta\nu_{i_\alpha n_\beta} \right) 
\end{eqnarray}
This equation is obtained assuming stationary currents and neglecting second-order terms. From left to right the right-hand side terms in the Ohm's law are: Ohmic, Hall, battery, and ambipolar term. The Ohm equation has similar form as the one for hydrogen plasma derived in e.g., \citet{Zaqarashvili2011} except that we use $\vec{u}_c$ instead of $\vec{u}_i$ and include more complete expressions for the coefficients that depend on the collisional frequencies.

The two-fluid formulation (\ref{eq:rhon}--\ref{eq:ei}) is applied when the difference in behavior between neutrals and ions is larger than between the neutrals/ions of different kind themselves. The forces acting on neutrals and charges are different since the magnetic Lorentz force only affects charges, and the neutrals feel its action exclusively via collisions. Thus, the two-fluid approach is based on a stronger coupling between charged particles than between charged and neutral particles  \cite{Zaqarashvili2011b, Zaqarashvili2011}. According to estimations of the magnetization factor of the solar atmosphere, the two-fluid approach is best valid at heights above $1000$ km.
Otherwise, when the collisional coupling of the plasma is strong enough, it is more convenient to use a single-fluid quasi-MHD approach i.e, including resistive terms that are not taken into account by the ideal MHD approximation. The conservation equations for the multi-species solar plasma become in this case:
\begin{eqnarray} \label{eq:continuity-single}
\frac{\partial \rho}{\partial t} + \vec{\nabla}\left(\rho\vec{u}\right) =  0
\end{eqnarray}
\begin{equation} \label{eq:momentum-single}
\frac{\partial (\rho\vec{u})}{\partial t} + \vec{\nabla}(\rho\vec{u} \otimes \vec{u})  = \vec{J}\times\vec{B} + \rho\vec{g}  - \vec{\nabla}{\bf\hat{p}} 
\end{equation}
\begin{eqnarray} \label{eq:energy-single}
\frac{\partial }{\partial t}\left(e + \frac{1}{2}\rho u^2 \right) & + & \vec{\nabla}  \left( \vec{u}\, ( e + \frac{1}{2}\rho u^2) +{\bf\hat{p}}\vec{u} + \vec{q}^{\,\, \prime} \right) + \vec{\nabla} \vec{F}_R  = \vec{J} \vec{E}^*   + \rho\vec{u} \vec{g}
\end{eqnarray}
where we use a similar notation as for the two-fluid formulation. This system of equations is closed by the generalized Ohm's law formulated for the single-fluid case, that reads as:

\begin{eqnarray} \label{eq:total_ohm}
\vec{E}^*=[\vec{E} + \vec{u}\times\vec{B}]= \frac{\rho_e}{(en_e)^2}\left(\sum_\alpha \nu_{ei_\alpha} + \sum_\beta \nu_{en_\beta} \right)\vec{J}   + \frac{1}{en_e}\ [\vec{J} \times \vec{B}]  -\frac{1}{en_e}  \vec{\nabla}{\bf\hat{p}}_e \\ \nonumber
-\frac{\xi_n^2}{\alpha_n}  [(\vec{J} \times \vec{B}) \times \vec{B}] +  \frac{\xi_n}{\alpha_n} [\vec{G} \times \vec{B}]
\end{eqnarray}
where $\vec{G}$ and $\alpha_n$ are given by
\begin{equation}
\vec{G} = \xi_n \vec{\nabla}{\bf\hat{p}}_{ie} - \xi_i \vec{\nabla}  {\bf\hat{p}}_n
\end{equation}
\begin{eqnarray}
\alpha_n &=& \sum_{\beta=1}^N\rho_e\nu_{en_\beta} + \sum_{\alpha=1}^N\sum_{\beta=1}^N\rho_i\nu_{i_\alpha n_\beta}
\end{eqnarray}
and $\xi_n=\rho_n/\rho$, $\xi_i =1-\xi_n$ are neutral and ion fractions. 

After applying Faraday's and Ampere's laws (neglecting Maxwell's displacement current) to the Ohm's equation formulated above, the induction equation is obtained for the evolution of the magnetic field:
\begin{eqnarray} \label{eq:induction-final}
\frac{\partial\vec{B}}{\partial t}  =  \vec{\nabla}\times \left[(\vec{u}\times\vec{B}) - \frac{\rho_e}{(en_e)^2}\left(\sum_\alpha \nu_{ei_\alpha} + \sum_\beta \nu_{en_\beta} \right)\vec{J} - \frac{1}{en_e}[\vec{J}\times\vec{B}] + \frac{1}{en_e}\vec{\nabla}{\bf\hat{p}}_e + \right. \nonumber \\ \left. + \frac{\xi_n^2}{\alpha_n} [[\vec{J}\times\vec{B}]\times\vec{B}]    - \frac{\xi_n}{\alpha_n}[\vec{G}\times\vec{B}] \right]
\end{eqnarray}

\noindent This equation closes the system of single-fluid equations. 

\section{Extended Mancha code}

The single-fluid equations \ref{eq:continuity-single}--\ref{eq:energy-single} and \ref{eq:induction-final} are solved by the extended version of the code \hbox{Mancha3D}. This is a non-ideal magnetohydrodynamic code with hyper diffusion algorithms and Cartesian grid written in Fortran 90. It solves the non-linear equations for perturbation in 2D, 2.5D and 3D, and requires an arbitrary magneto-hydrostatic equilibrium to be explicitly removed from the equations. Spatial discretization is based on a six-order centre-difference scheme. The numerical solution of the system is advanced in time using an explicit fourth-order Runge-Kutta scheme. The code is fully MPI-parallelized using the distributed memory concept, and allowing full arbitrary 3D domain decomposition. A description of an early purely MHD version of this code can be found in \cite{Felipe+etal2010}. Compared to the version described in that paper, the new code includes the non-ideal terms derived from the generalized Ohm's law (Eq. \ref{eq:total_ohm}), and the module for the calculation of non-ideal equation of state and the RT module that solves the Radiative Transfer equation (Eq. \ref{eq:transfer}).

Order of magnitude estimates of the importance of the different terms in the generated single-fluid Ohm's law (Eq.\ref{eq:total_ohm}) done in \cite{Khomenko+etal2014b} for various atmospheric models show that only ambipolar, Hall and (potentially) battery terms can reach significant values in the photosphere and chromosphere of the Sun. Therefore, in the current version of the code we only retain these three terms.  This feature is partially described in \cite{Khomenko+Collados2012, Khomenko+etal2014}.

The native set of variables used in the code is $\{ \rho, \rho\vec{u}, e, \vec{B}  \}$. The pressure, $p$, is assumed to be a scalar and is obtained from $e$ through the implementation of realistic Equation of State (EOS) for the solar mixture that takes into account the effects on ionization. This is done by assuming a non-ideal gas in thermodynamical equilibrium in the convection zone, and an ideal partially ionized gas of a mixture including molecules in the photosphere, and is implemented as interpolations of lookup tables. An ideal gas with parametrized thermodynamical quantities is used as the first approximation elsewhere. We compute additionally the tables for electron pressure, so that the ionization fraction is obtained self-consistently with the rest of thermodynamical variables. We assume instantaneous ionization balance. The EOS implemented in that way provides the required precision while enormously reducing the computing time.

The newly implemented RT module solves the non-gray Radiative Transfer equation (Eq. \ref{eq:transfer}) assuming Local Thermodynamic Equilibrium (LTE). The wavelength dependence of the emission and absorption coefficients $j_{\nu}$ and $k_{\nu}$ is discretized by the opacity binning method \cite{Nordlund1982}. The angle discretization is set prior to the code execution. Both discretizations are performed by sets of routines accompanying the code. The formal solver used in the RT module is based on the short-characteristics method \cite{Olson+Kunasz1987}. The module is fully parallelized using MPI. The LTE approximation limits the application of the RT module to simulations of the top of the convective zone and the photosphere. In the optically thin corona, the code uses Newtonian cooling approximation. 

The code has passed standard tests on numerical performance and speed, with scaling tests in up to several thousand cores, showing its reasonable efficiency. The code is modular, includes conditional compilation and a modern input/output file format (HDF5). An extended 2/2.5D version of the code uses Adaptive Mesh Refinement (AMR) implemented via the PARAMESH library. 

\begin{figure}[!]
\center
\includegraphics[width=16cm]{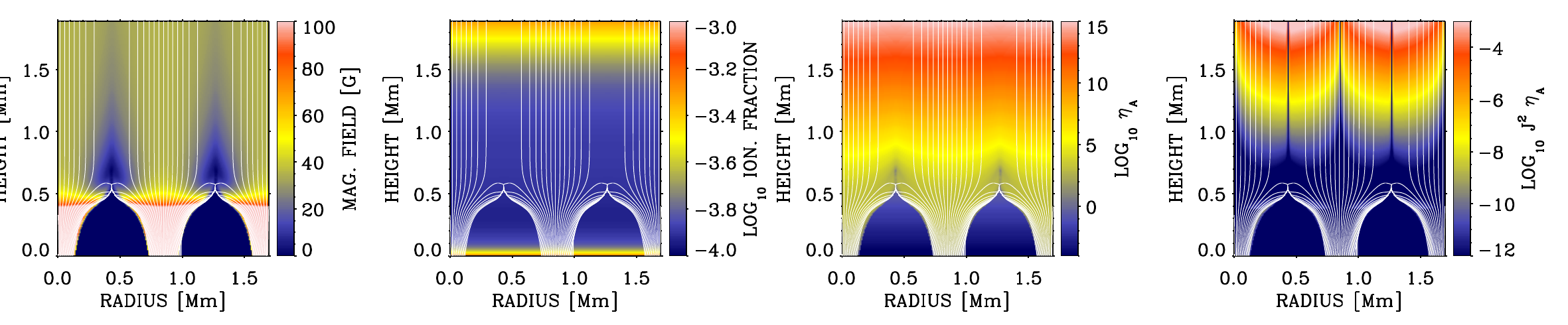} 
\caption{From left to right: initial distribution of the magnetic field; ion fraction $\xi_i$; ambipolar diffusion coefficient $\eta_A=\xi_n^2|B|^2/\alpha_n\mu_0$ (in m$^2$ s$^{-1}$); and the quantity $J^2 \eta_A$ in a flux tube model.} \label{fig:holmul}
\end{figure}

\begin{figure}[!]
\center
\includegraphics[width=7cm]{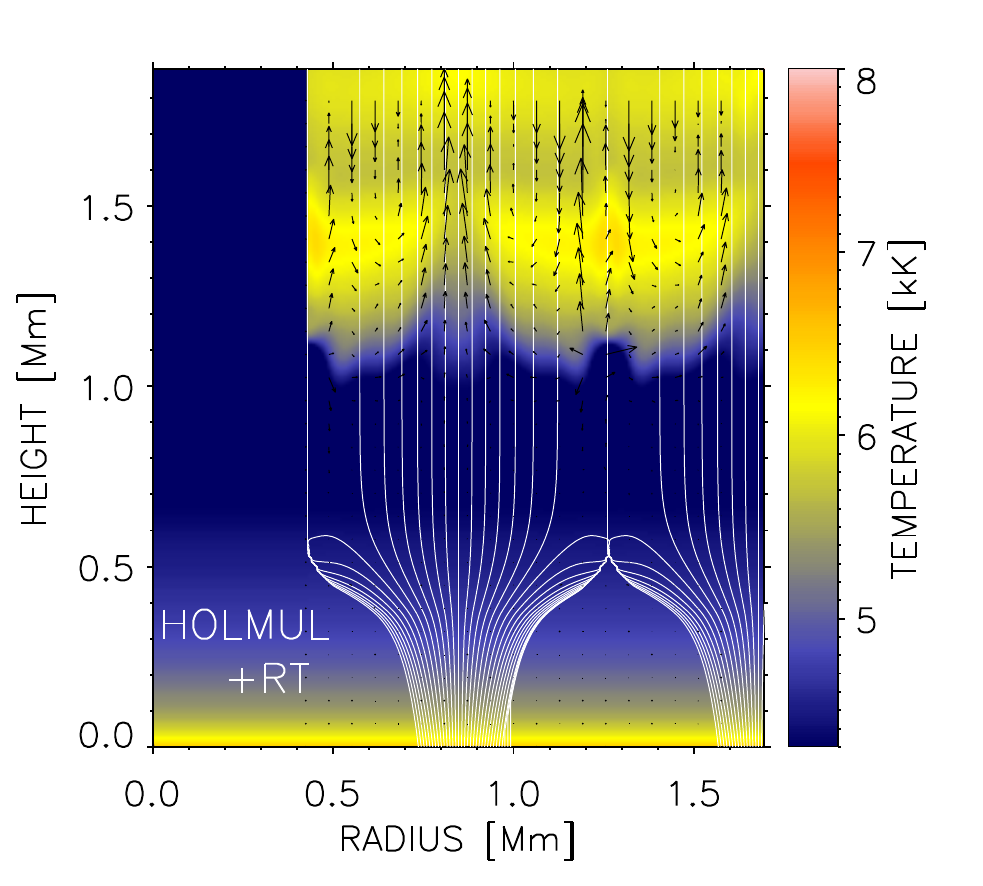} 
\includegraphics[width=7cm]{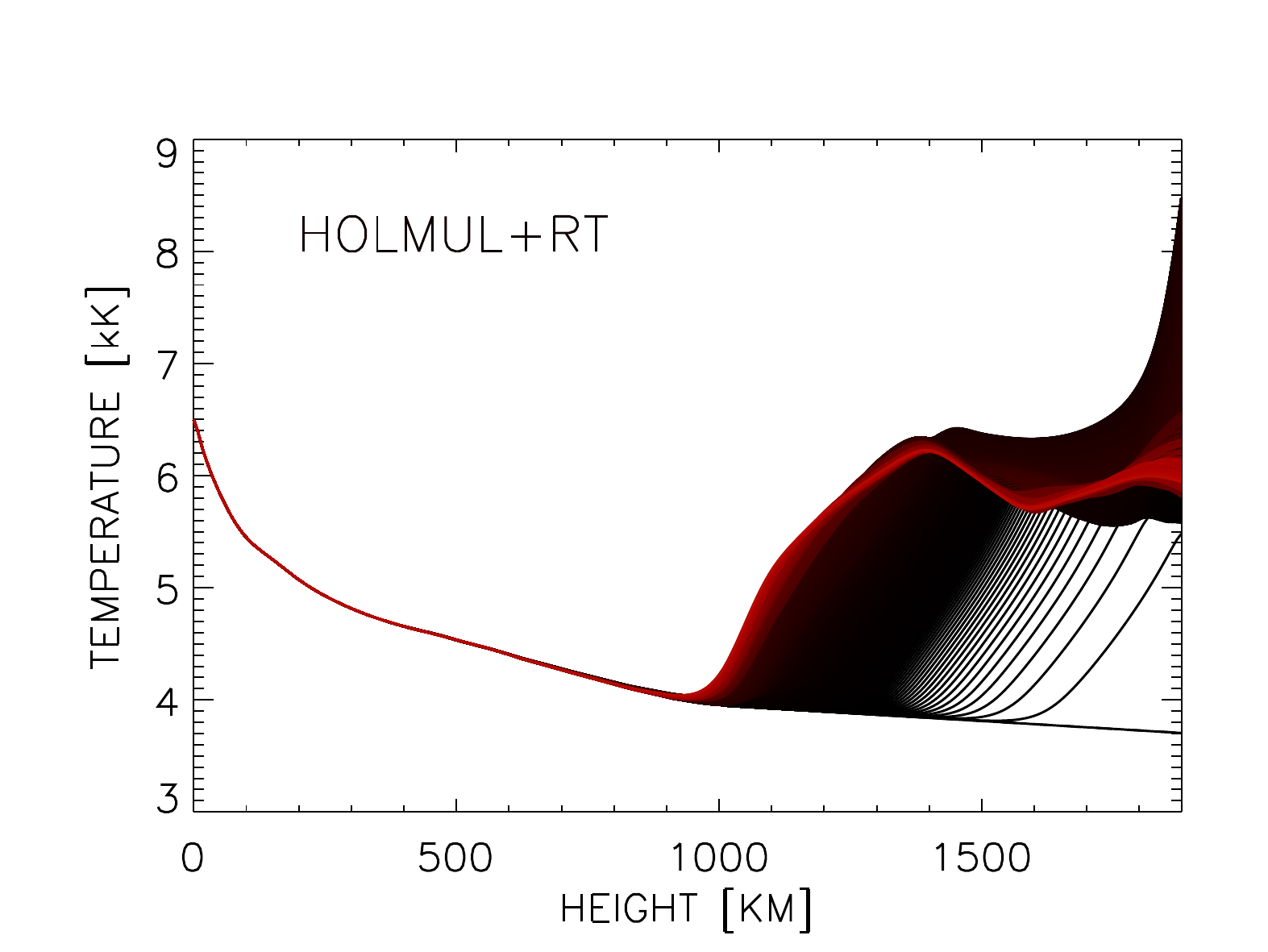} 
\caption{Simulations of chromospheric heating by ambipolar diffusion. Left: simulation snapshot at 800 s. The background color is temperature, white lines are magnetic field lines, arrows show the velocity field. For better visual comparison the temperature structure from 0 to 0.42 Mm is kept unchanged from its initial value. Right: temperature as a function of height at horizontal position 0.6 Mm. Different lines are separated 0.5 sec in time; progressively more red colors indicate larger times till 800 sec since the start of the simulation.} \label{fig:holmul-temp}
\end{figure}

\subsection{Chromospheric heating due to partial ionization effects}

Dissipation of currents due to the action of the ambipolar term provides an efficient source of heat. We performed 2.5D simulation of the chromospheric heating in an idealized atmospheric model composed of a set of flux tubes. We used a 2nd-order thin magnetic flux tube model constructed after  \cite{Pneuman+etal1986}, see Figure \ref{fig:holmul}.  The model represents a horizontally infinite series of flux tubes that merge at some height in the chromosphere. This magnetic field configuration is non force-free. In order to evaluate the efficiency of the heating mechanism by ambipolar diffusion, we intentionally selected the temperature stratification of the flux tubes to be given by a model HOLMUL \cite{holmul}, monotonically decreasing with height dictated by the conditions of radiative equilibrium. The ionization fraction in this model does not exceed $10^{-3}$, even in the chromosphere.  The ambipolar diffusion coefficient, defined as $\eta_A=\xi_n^2|B|^2/\alpha_n\mu_0$ is orders of magnitude larger than the Ohmic one, reaching values of 10$^{15}$ m$^2$ s$^{-1}$  at a height of 2 Mm. Such values of $\eta_A$ imply important current dissipation on very short time scales. 

The initial model is in MHS equilibrium and does not evolve without external perturbation. As a perturbation we use ambipolar term in the induction (\ref{eq:induction-final}) and energy (\ref{eq:energy-single}) equations. Thanks to the Joule heating in the energy equation, $\vec{J}\vec{E}$, the magnetic energy is efficiently converted into thermal energy, producing heat. This heat is balanced by the radiative cooling term, $\vec{\nabla} \vec{F}_R$. The left panel of Figure~\ref{fig:holmul-temp} shows the simulation snapshot 800 s after the introduction of the perturbation. The right panel of this figure shows the height dependence of the temperature at a fixed horizontal position inside the flux tube for different time moments. The most important heating is achieved at the upper part of the domain, close to the tube borders, since the term responsible for the heating is orders of magnitude larger at these locations (right panel of Fig. \ref{fig:holmul}).

At the first few seconds of simulation, the chromospheric temperature significantly increases reaching up to 8 kK at the upper part of the domain, i.e. about 4 kK above its initial value. The time scales associated to the Joule heating are very short, of the order of seconds, similar to those of radiative cooling. Due to the balance between these two processes, after the initial heating, we observe damped temperature oscillations while converging to some constant value about 6 kK, forming hot chromosphere. Thus, we conclude that the Joule heating and radiative cooling terms in the energy equation can balance each other and lead to the chromospheric temperature rise. In a more complex situation, magneto convection simulations performed in \cite{MartinezSykora+etal2012} show that the action of the ambipolar diffusion allows to remove artificially cool bubbles produced by adiabatic plasma expansion in the chromosphere and to provide a more realistic model of its temperature structure. Therefore it is important to include ambipolar diffusion in further models of the chromosphere.

\begin{figure*}[!t]
\center
\includegraphics[width=16cm]{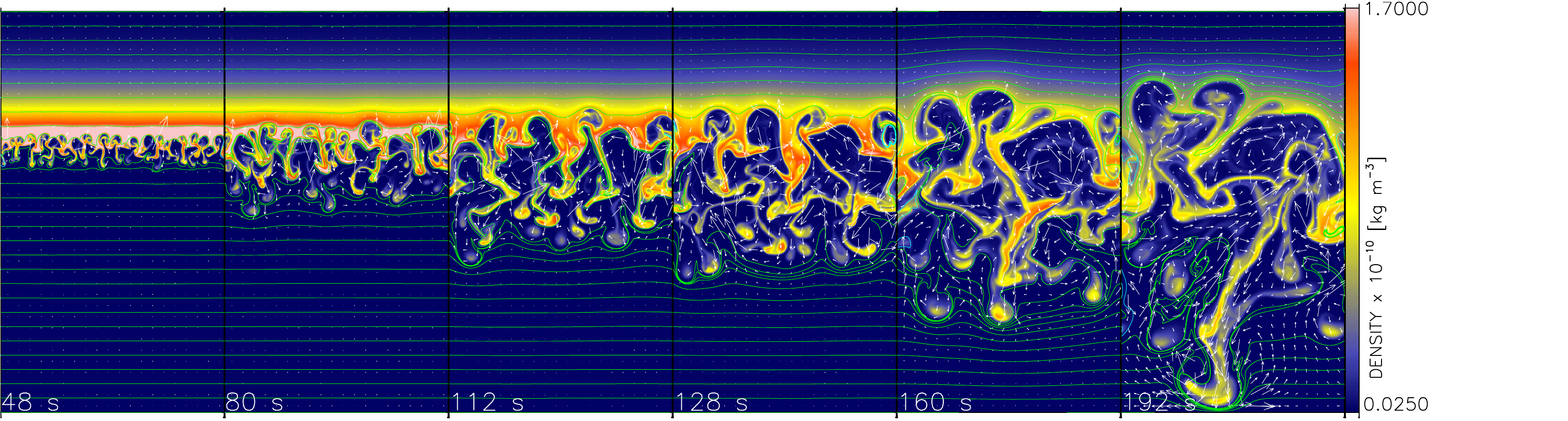}
\includegraphics[width=16cm]{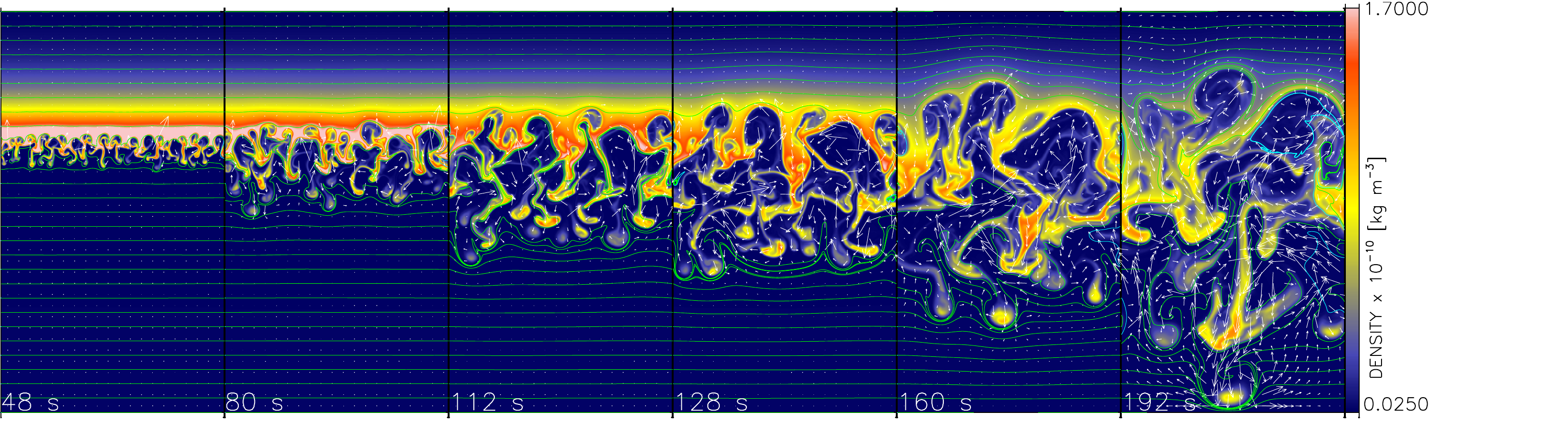}
\caption{{\footnotesize Time evolution of density in the simulations of Rayleigh-Taylor instability including ambipolar term (top) and without this term (bottom). The size of each snapshot is 1$\times$1.8 Mm, the elapsed time is given at the bottom of each panel. Green lines are the projections of the magnetic field lines into the perturbation plane, the velocity field is indicated by arrows. } }\label{fig:rho}
\end{figure*}

\begin{figure*}
\center
\includegraphics[width=5cm]{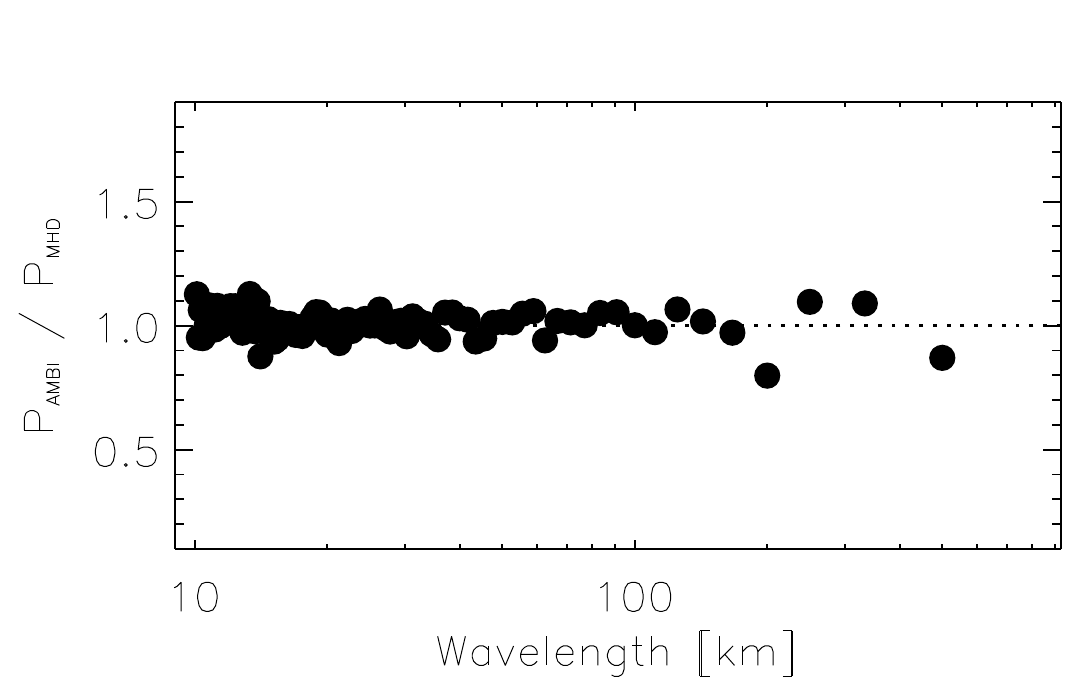}
\includegraphics[width=5cm]{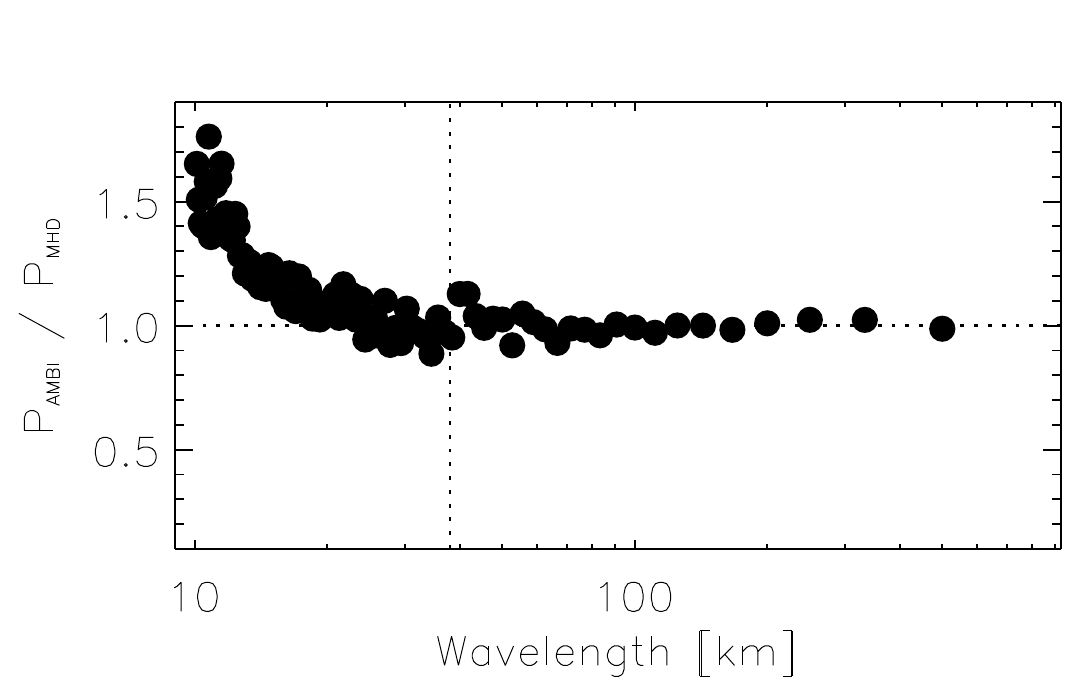}
\includegraphics[width=5cm]{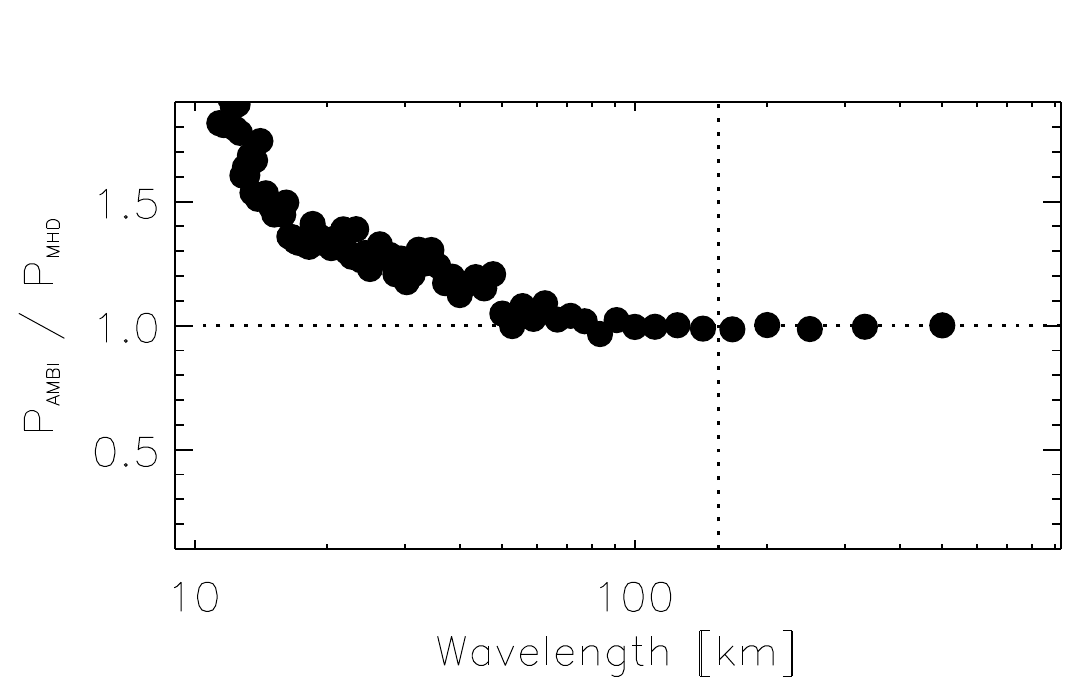}
\caption{Spectrum of the relative power of the ``ambipolar'' to ``mhd'' simulations of RTI with magnetic field normal to the perturbation plain ($\theta=90$\degree, top) and skewed from the plain at  $\theta=89$\degree (middle) and $\theta=88$\degree (bottom). In the last two cases, the vertical dotted lines mark the cut-off wavelength of the instability, $\lambda_c=38$ and $155$ km.}
\label{fig:scales}
\end{figure*}

\subsection{Rayleigh-Taylor instability in solar prominences}

Plasma of solar prominences is yet another environment where the manifestation of partial ionization effects can be important. Typical values of temperature, derived from the Doppler width of spectral lines observed in prominences, are of the order of 8-9 kK \cite{Stellmacher+etal2003, Labrosse+etal2010}. The ionization fraction is less known, but the values provided in the literature vary in the range $\xi_i$ =0.2--0.7 \cite{Labrosse+etal2010}. The prominence-corona transition region (PCTR) is thought to be subject to tangential instabilities, as the Kelvin-Helmholtz instability (KHI) or the Rayleigh-Taylor instability (RTI) \citep{Berger+etal2008, Schmieder+etal2010, Heinzel+etal2008, Ryutova+etal2010}. The linear theory has demonstrated that both instabilities are affected by the presence of neutrals in the prominence plasma \cite{Soler+etal2012, Diaz+etal2012, Diaz+etal2013}.

We performed 2.5D simulations of the non-linear phase of the RTI at the PCTR considering partial ionization of the  prominence plasma in the single-fluid formulation. As initial setup we used a purely hydrostatic stratification of pressure and density for a given constant temperature in the prominence (5 kK) and coronal (400 kK) parts of the domain. A constant horizontal magnetic field of 10 G was set over the whole domain, and its orientation with respect to the perturbation plane varied from one simulation to another. For these parameters, the neutral fraction in the prominence makes $\xi_n=0.9$ and the ambipolar diffusion coefficient is $\eta_A=2.3\times 10^8$ m$^2$ s$^{-1}$. The instability is initiated by a multi-mode perturbation of the interface position. For each orientation of the magnetic field, we compare simulations with and without ambipolar term, see \cite{Khomenko+etal2014}. Figure \ref{fig:rho} shows a time series of simulation snapshots illustrating the development of the instability for the case with ambipolar term on (top) and off (bottom). The flows start the same, but then develop differences on small scales, since the ambipolar term introduces additional diffusion acting perpendicular to the magnetic field. These slight differences in the particular form of the turbulent flows can be observed at the end of the simulation. The ambipolar term is larger in the regions with low density, i.e. regions with, generally, low momentum $\rho \vec{u}$. Therefore, the impact of such regions into the overall flow dynamics is not large. A similar behavior is observed in the simulations of multi-fluid turbulence in molecular clouds \cite{Downes+Sullivan2011, Downes2012}. However, statistically, the flows in both cases are different, as is shown in Figure \ref{fig:scales}. In agreement with the linear theory, non-linear simulations demonstrate that the introduction of the ambipolar diffusion removes the cut-off wavelength for the growth rate of the instability and allows the small scales to develop. As the non-linear development of the instability is such that small scales merge and give rise to larger scales \citep{Jun+etal1995b, Stone+Gardiner2007a,  Hillier+etal2012a}, the larger growth rate of small scales leads to larger extreme velocities in the ambipolar case \cite{Khomenko+etal2014}. We also find up to 30\% larger temperatures at the PCTR in the simulations with ambipolar term, as a result of the Joule dissipation. A significant drift momentum, defined as $\vec{p}_D=\sqrt(\rho_i\rho_n)(\vec{u}_i-\vec{u}_n)$ is present at the PCTR \cite{Khomenko+etal2014}. 

\section{Observational manifestations of ion-neutral effects}

Direct observational confirmation of the uncoupled behavior of neutral and ionized species in the solar atmosphere is still missing. Order of magnitude calculations show that the scales at which ion-neutral effects become important are rather small. At maximum, spatial scales reach few kilometers, and temporal scales reach fraction of seconds, depending on the height in the atmosphere and on the magnetic field strength \cite{Khomenko+etal2014b}. Direct observation of such small scales is not possible with current observational facilities. Nevertheless, if the differences in the behavior of ions and neutrals are sufficiently large in a small area inside the resolution element, it may be possible that their traces remain in the integrated signal 
even at the typical resolution of observations. This gives a chance of their detection in a specially dedicated observational campaign.  Below we describe the preliminary results from two  such campaigns. In one case, we target a prominence plasma. In the second case, the target was the Evershed flow in sunspots.

\begin{figure*}[!]
\center
\includegraphics[width=10cm]{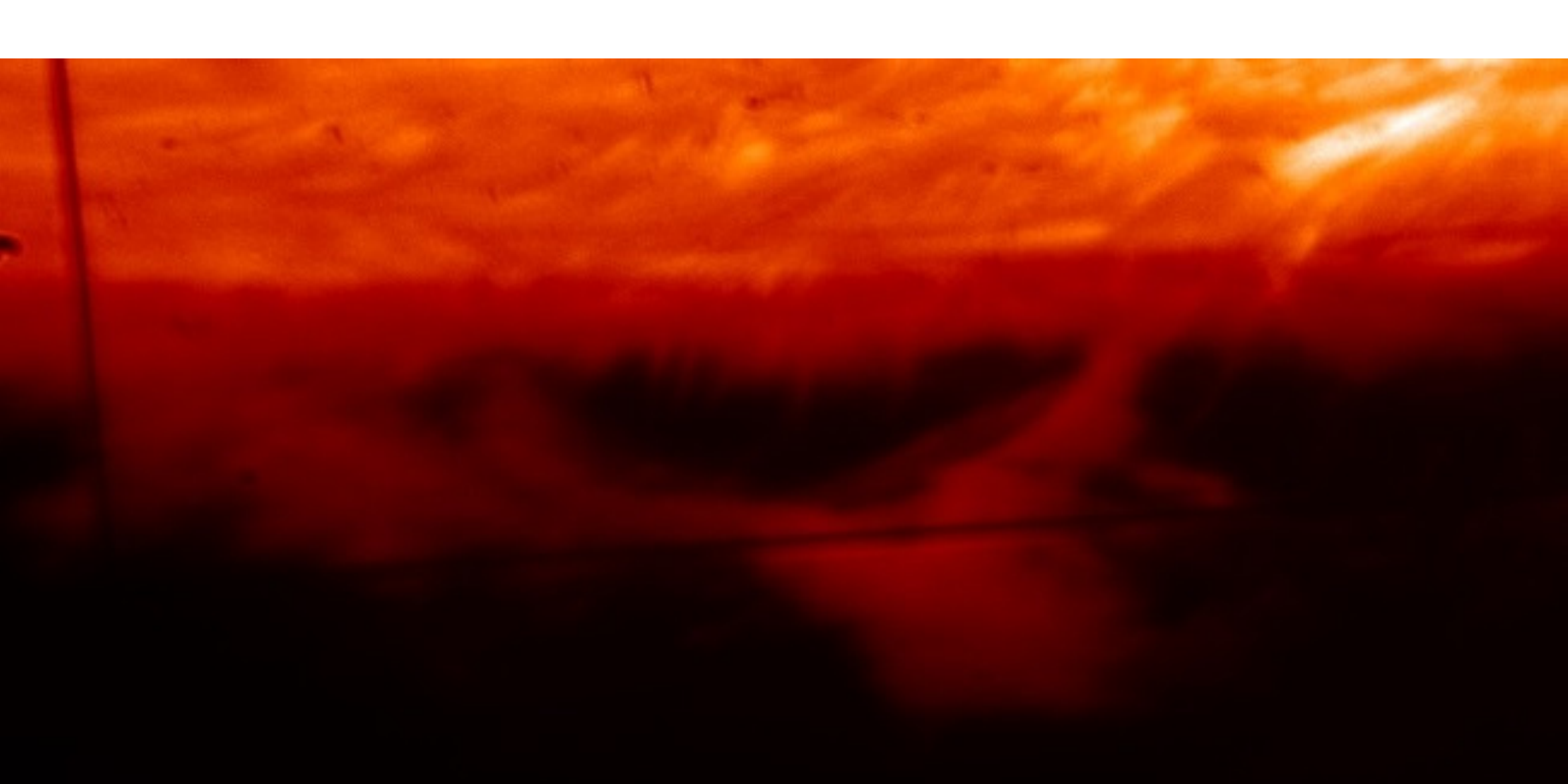}
\caption{Slit-jaw image in \Halpha\ of the prominence observed on 11th of September, 2012. Black nearly horizontal line shows the location of the slit in one of the scans.}
\label{fig:sj}
\end{figure*}

\begin{figure*}[!]
\center
\includegraphics[width=5.0cm]{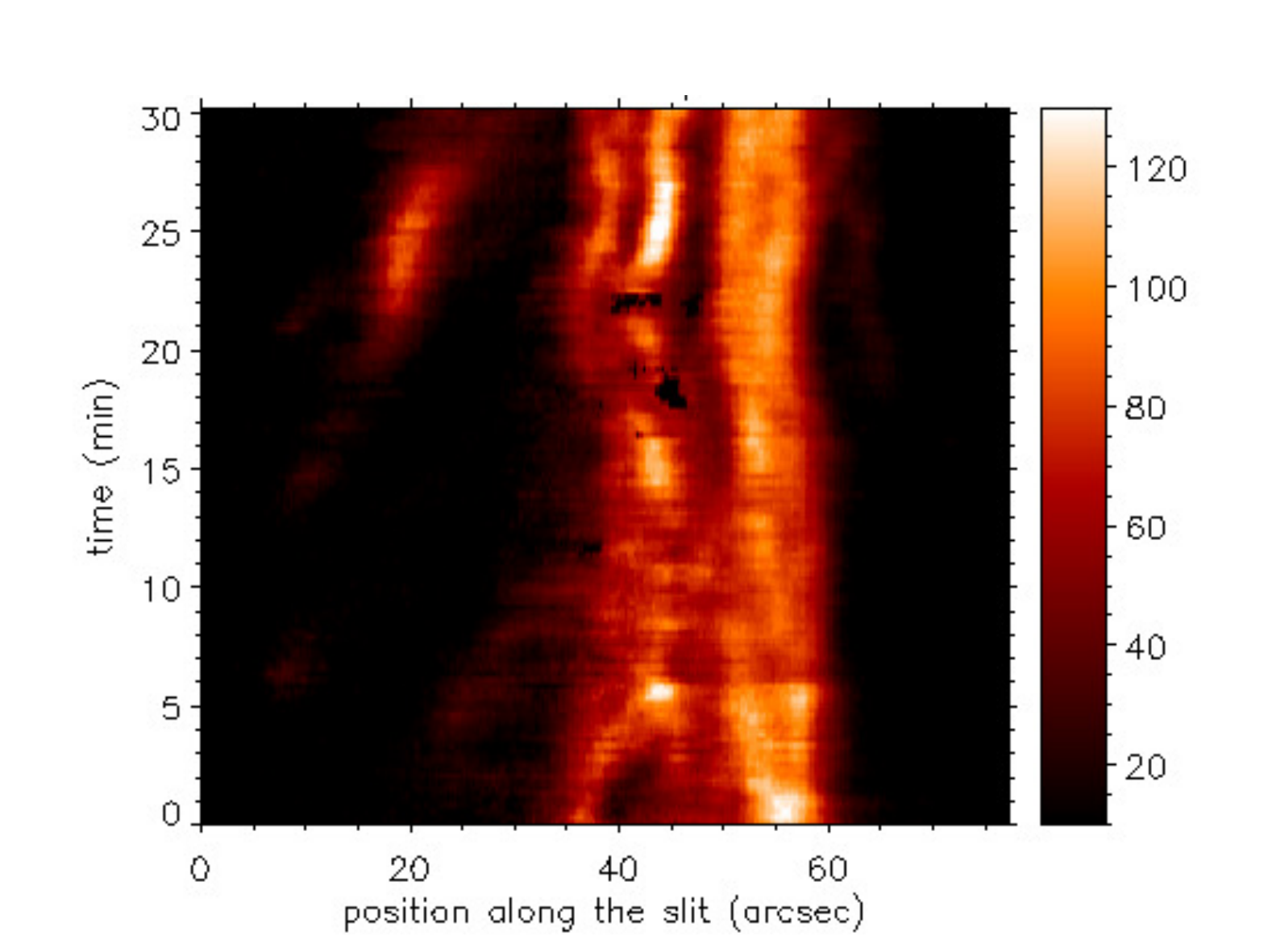}
\includegraphics[width=5.0cm]{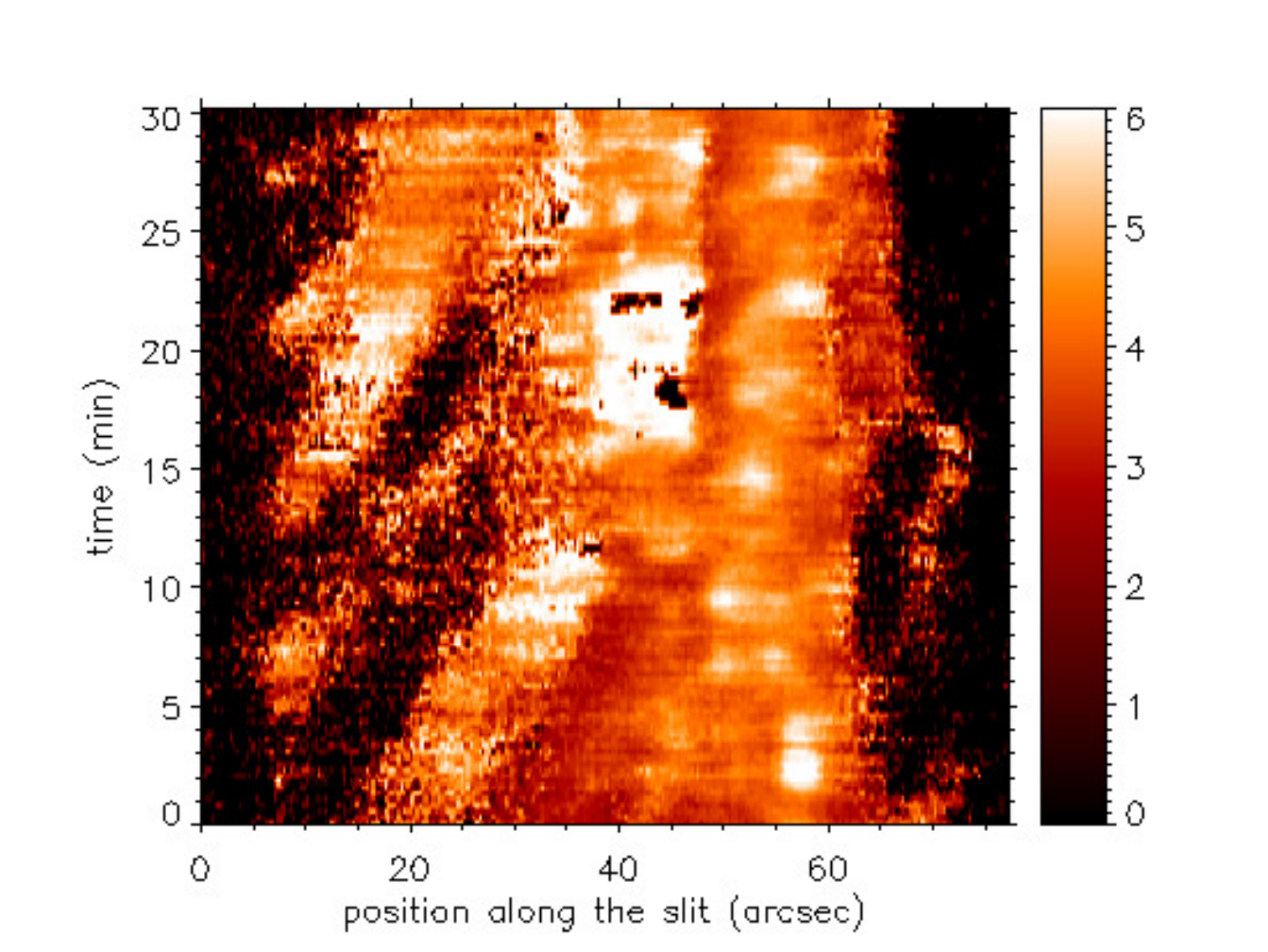}
\includegraphics[width=5.0cm]{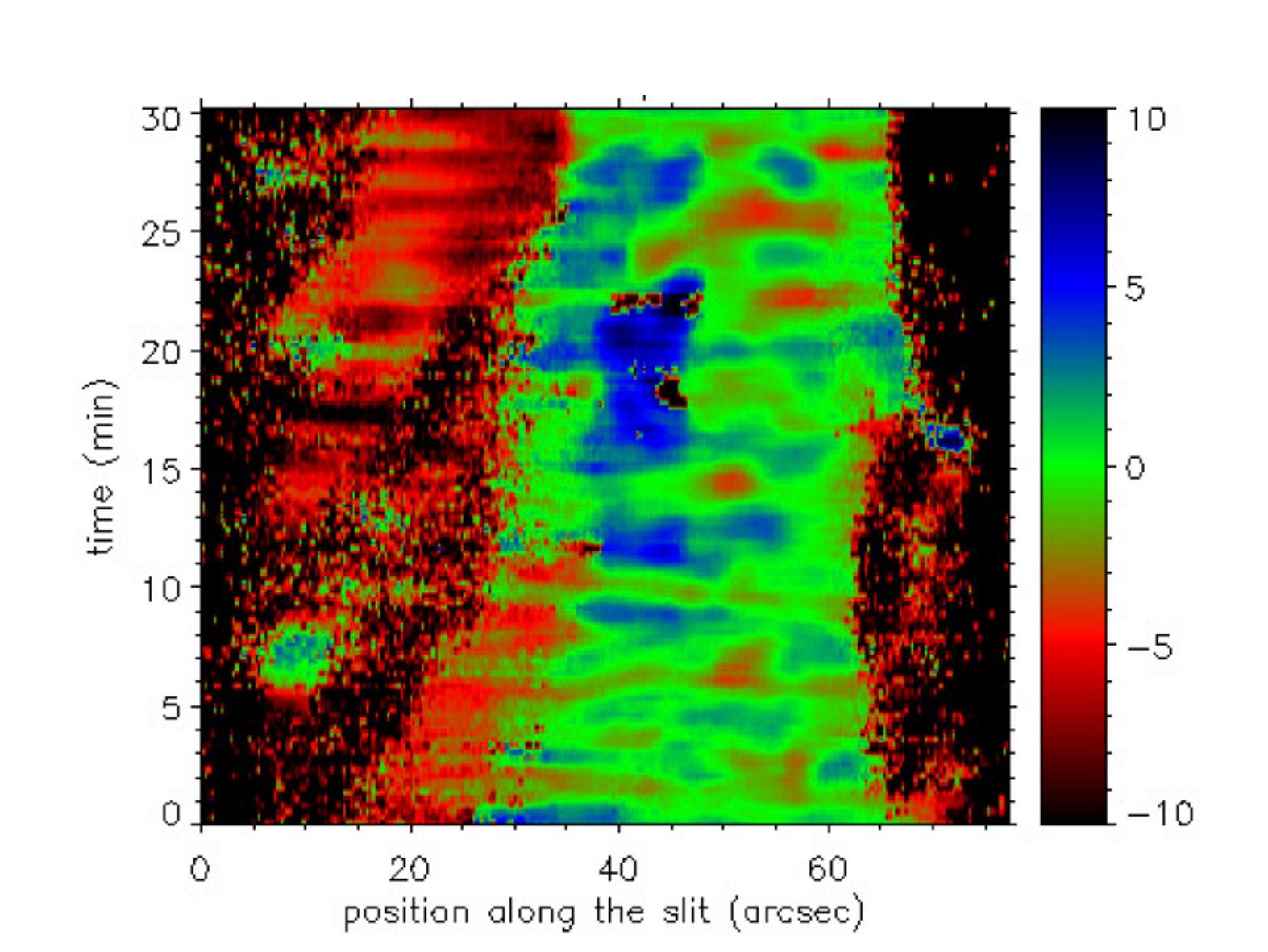}
\includegraphics[width=5.0cm]{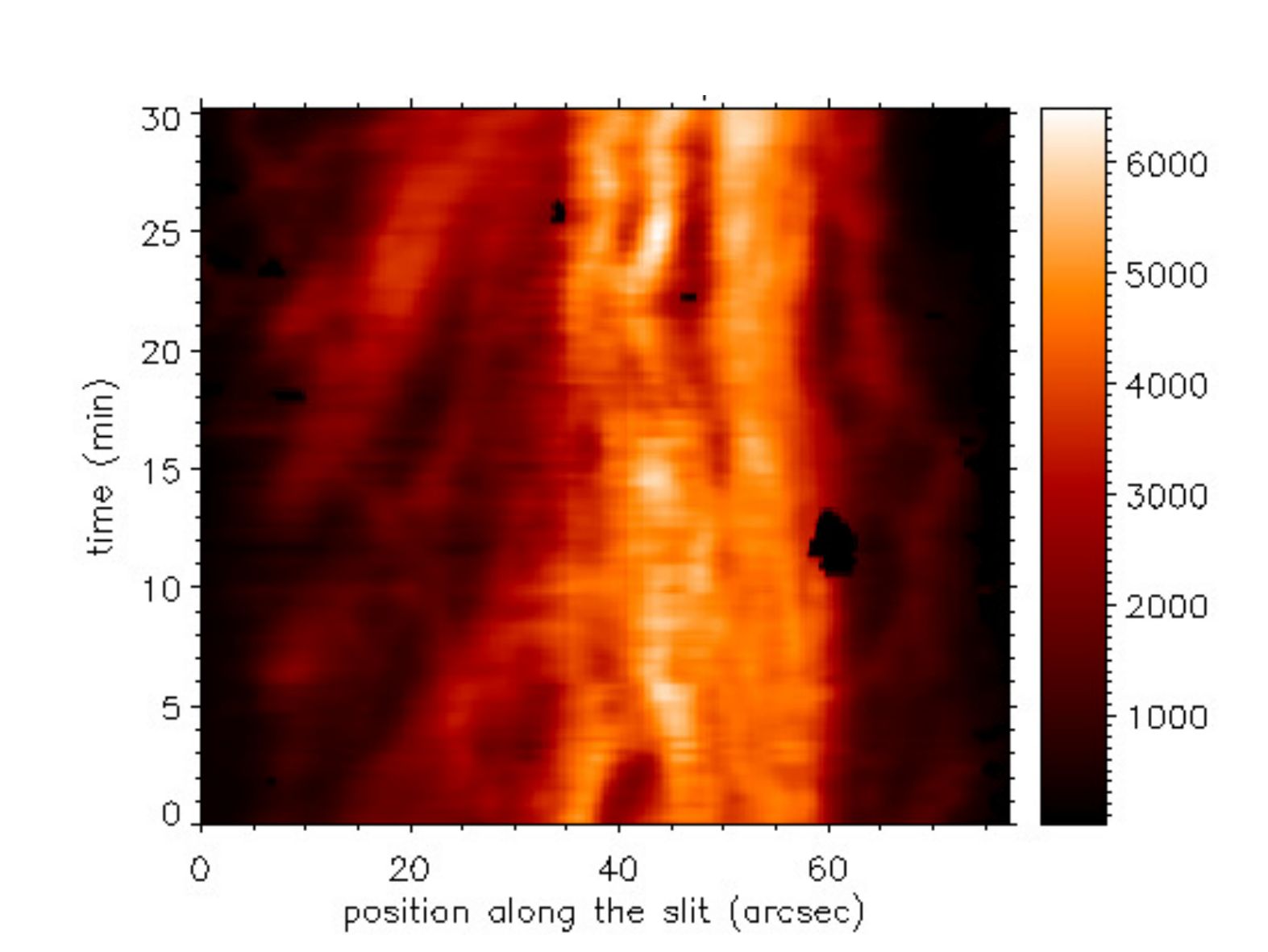}
\includegraphics[width=5.0cm]{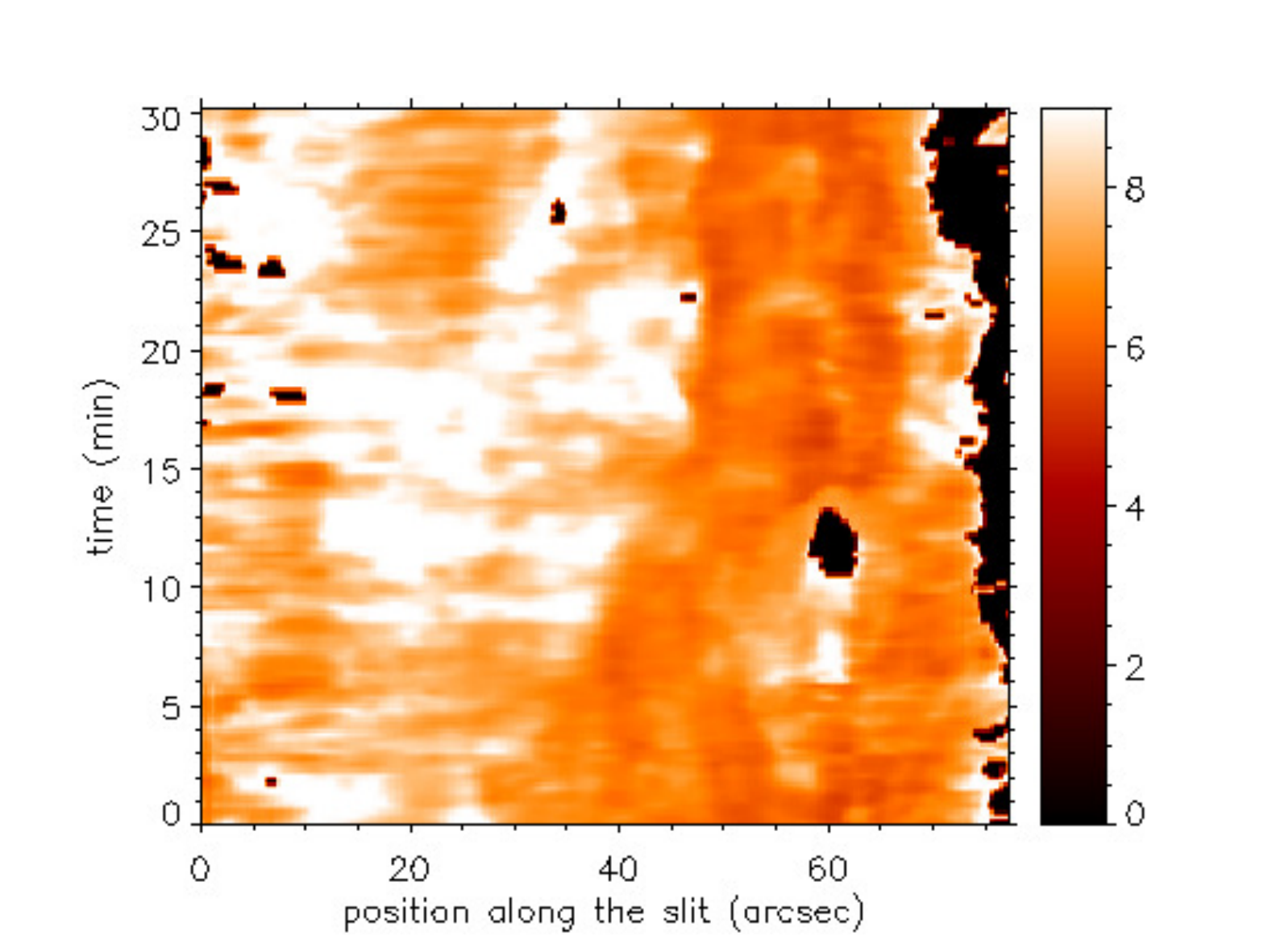}
\includegraphics[width=5.0cm]{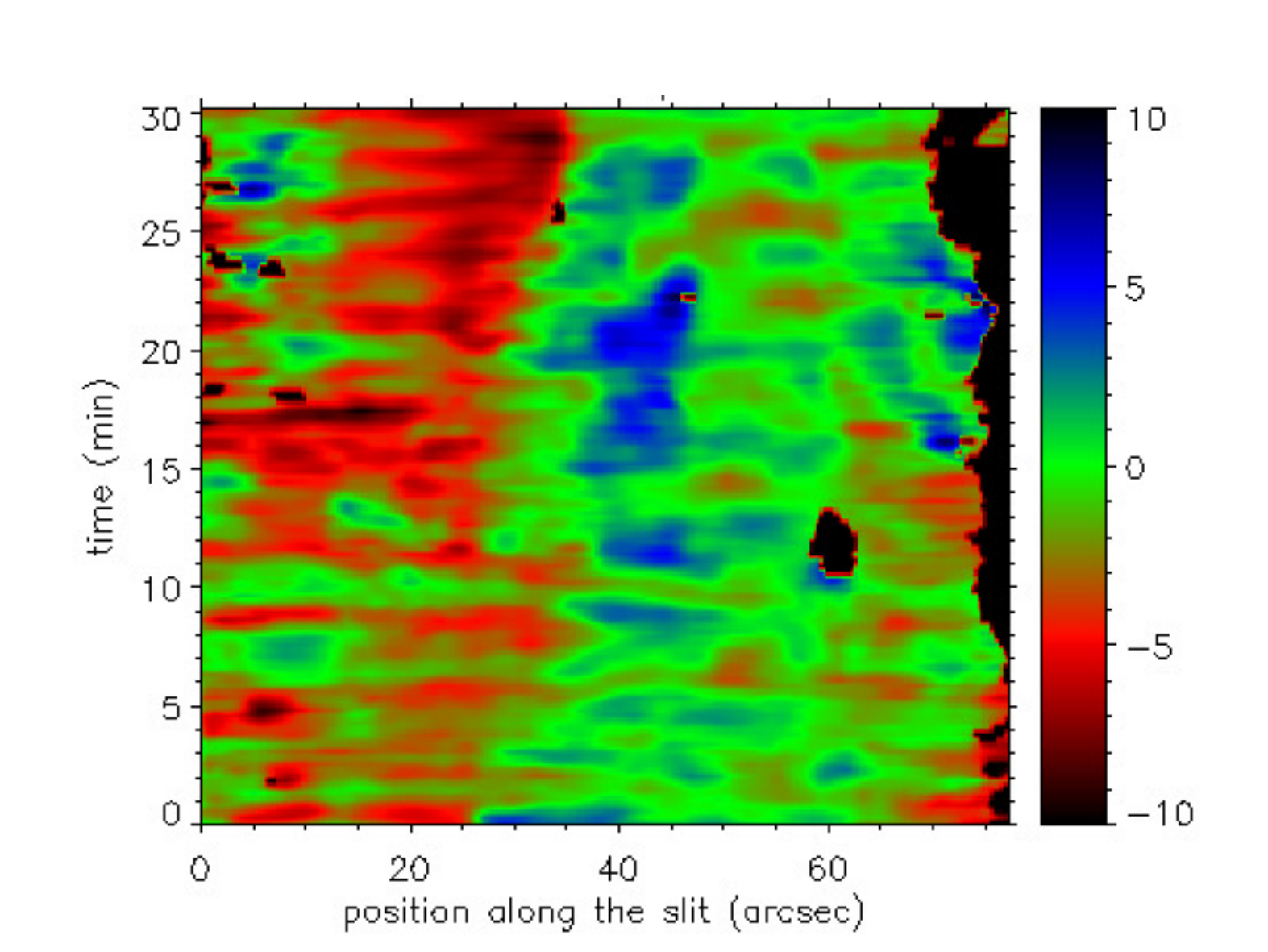}
\caption{Top: amplitudes (left), widths (middle) and displacements (right) of the \CaII\ 4582 \AA\ line as functions of the position along the slit and time. Displacements are given in \kms. Bottom: same for the \HeI\ 10830 \AA\ line.}
\label{fig:fit}
\end{figure*}

\subsection{Prominences}

The observations were done on 11th of September, 2012, using the Vacuum Tower Telescope (VTT) at the Observatorio del Teide on Tenerife. We targeted a prominence located at the limb close to the active region NOAA 11564 that was undergoing evolution during the observations.  A slit-jaw image of the observed prominence is shown in Figure \ref{fig:sj}. The data represent a time series of 10-steps scans (0.14\arcsec\ per step) of a small part of the prominence. It took 1.5 sec per scan position, a fixed position was repeatedly measured every 15 sec. We observed \CaII\ 8542 \AA\ and \HeI\ 10830 \AA\ spectral lines, by simultaneously using the Tenerife Infrared Polarimeter II (TIP \cite{Collados2007}) and the spectrograph of the VTT. Our primary objective was fast scanning in order to follow the rapid changes in the dynamics of the prominence. Therefore, no polarimetry was performed.  The data acquisition started at 08:49 UT for almost 30 min, with the $r_0$ parameter staying around 10 cm. Standard reduction of the data was performed including the correction for the differential refraction for spatial alignment of the spectra taken at the two different wavelengths. The detailed description of the data and their analysis is given in \cite{Diaz+etal2015}.

\begin{figure*}[!]
\center
\includegraphics[width=7.0cm]{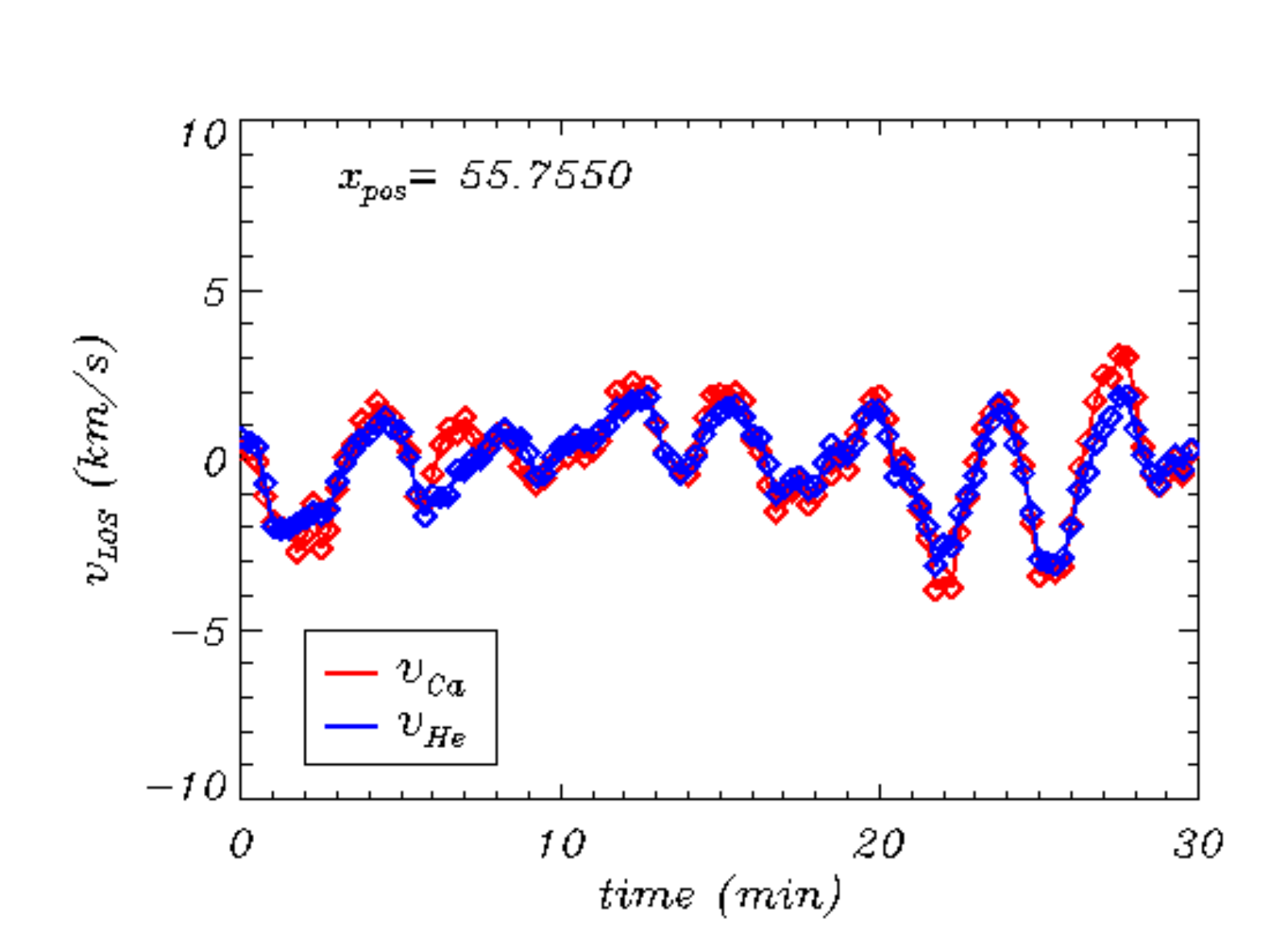}
\includegraphics[width=7.0cm]{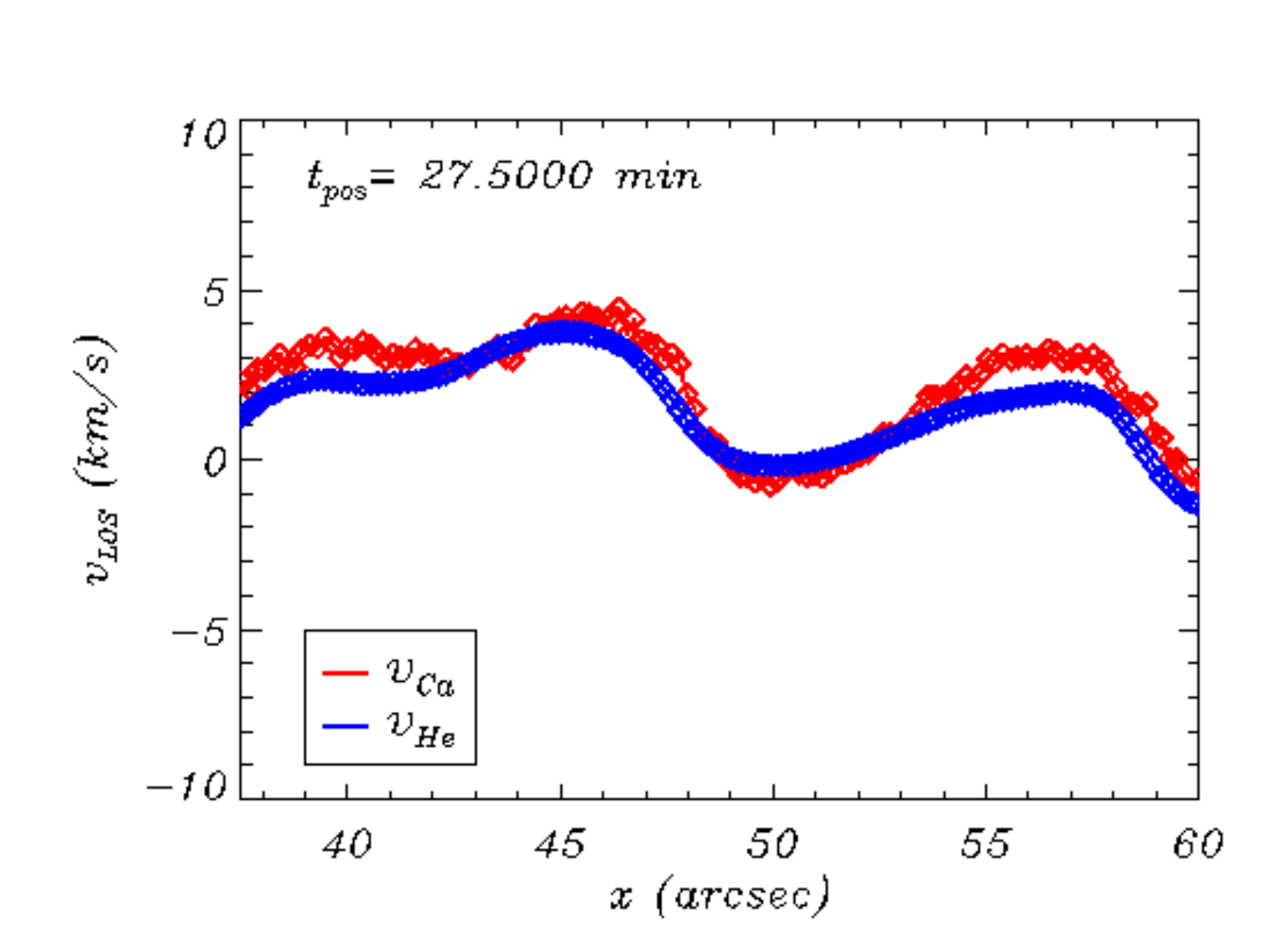}
\caption{Left: velocity measured from \CaII\ line (red) and from \HeI\ line (blue) as a function of time at a fixed location. Notice the oscillatory behavior and  slightly larger amplitudes in Ca at the velocity extremes. Right: velocities in both lines along the slit for a fixed time. Notice the lag between both velocities at the locations with larger gradients.}
\label{fig:vel-cut}
\end{figure*}

The scientific analysis included a gaussian fit to the observed emission spectral lines, from which we extracted their amplitudes, widths and displacements \cite{Diaz+etal2015}. The results of this fit are given in Figure \ref{fig:fit}. The upper raw of this figure gives the results for the \CaII\ line while the bottom one is for the \HeI\ line.  Due to the low sensitivity of the CCD at the wavelength of \CaII\ line the data for this line are much noisier than for \HeI. Therefore, there are locations where we could not perform any fit of the \CaII\ line, or the quality of the fit was not adequate. In most of the locations both lines show simple shapes that can be interpreted as due to a single-component atmosphere along the line of sight. However, at some points the profiles show a clear two-component structure due to the appearance of the surges in the line of sight. In the preliminary analysis detailed here, we did not give any special treatment to these profiles. Figure \ref{fig:fit} shows that, outside of the locations where the \CaII\ line could not be fitted, the distribution of the amplitudes and widths of both lines in time and along the slit is rather similar. To first order, we take this as a confirmation that both lines form at the same location in the prominence and therefore measure the dynamics of the same plasma. It can be observed at the right panels of Fig. \ref{fig:fit} that indeed, the velocities derived from both lines are very similar. 
A more detailed look at the velocities is given in Figure \ref{fig:vel-cut}. It shows two selected examples of the time dependence at a fixed location (left) and spatial dependence at a fixed time (right). It can be observed that, while the velocities in \CaII\ and \HeI\ are very similar, differences appear at the locations of extreme velocities (left panel) or at the locations of more rapid changes in space where the velocity gradients are stronger (right panel). Having in mind the preliminary nature of our analysis, it may indicate that velocities of neutral and ionised atoms in prominences shows differences, partially attributing this to ion-neutral effects.

\begin{center}
\begin{table}
\begin{center}
\caption{Parameters of the spectral lines in sunspot observations. Atomic data are taken from \cite{Gurtovenko+Kostik1989}. Formation heights are calculated in the penumbra models C and D from \cite{Socas-Navarro2007}.} 
\begin{tabular}{cccccccc}
\hline Elm Ion & $\lambda$ (\AA\ )  &  EPL (eV) & $\log gf$ & $g_{\rm eff}$ & $H_C$(km) & $H_D$(km) & I \\ \hline
\FeII\ & 4576.3  & 2.84  & $-2.976$ & 1.18  &  449  & 293  &  0.69\\
\FeI\  & 4574.2  & 3.21  & $-2.500$ & 1.90  &  280  & 276  &  0.48\\
\hline
\FeII\ &  4656.9 &  2.89 & $-3.643$ &  1.67  & 162 & 171 &  0.41\\
\FeI\  &  4657.6 &  2.85 & $-2.900$ &  1.40  & 277 & 274 &  0.41\\
\hline
\FeI\  & 5196.1  & 4.26 & $-0.493$ &  1.09  & 336  & 350 &  0.69\\
\FeII\ & 5197.6  & 3.23 & $-2.348$ &  0.67  & 539  & 328 &  0.72\\
\FeI\  & 5197.9  & 4.30 & $-1.640$ &  0.36  & 216  & 194 & 0.39\\
\FeI\  & 5198.7  & 2.22 & $-2.135$ &  1.50  & 398  & 494 & 0.79\\
\hline
\FeII\ & 5234.6 & 3.22 & $-2.279$ &  0.87 &  553   & 331 & 0.72\\
\FeI\  & 5236.2 & 4.19 & $-1.497$ &  0.39 &  235   & 244  & 0.38\\
\hline
\FeII\ & 6516.1 &  2.89 & $-3.432$ &   1.07 &  281 & 250 & 0.46\\
\FeI\  & 6518.4 &  2.83 & $-2.460$ &   1.15  & 327 & 351 & 0.51\\
\hline
\end{tabular}
\end{center}
\end{table}
\end{center}

\begin{figure*}[!]
\center
\includegraphics[width=16.0cm]{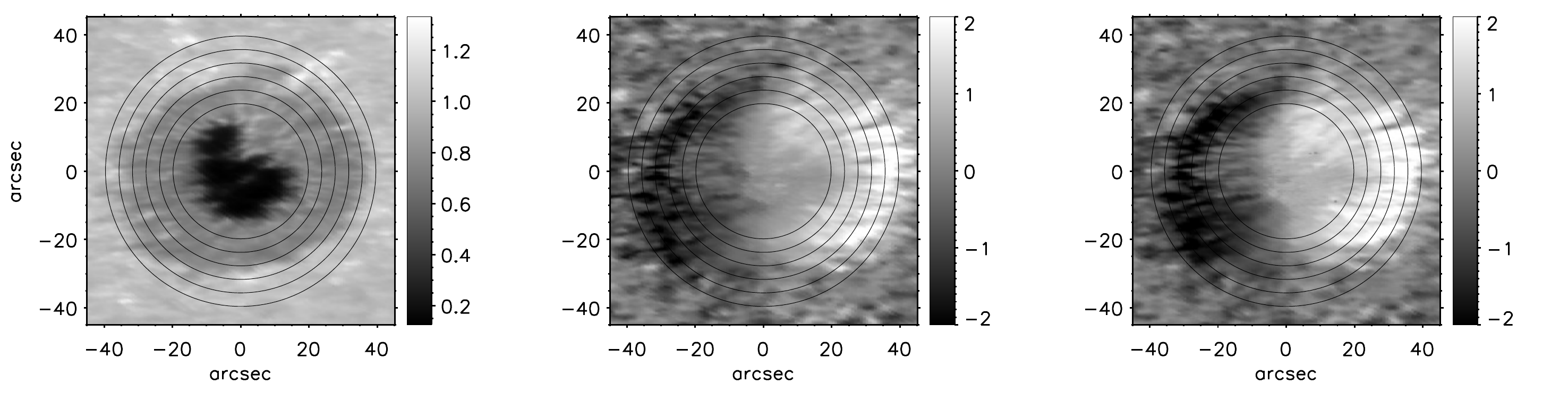}
\caption{Continuum image and velocities of the \FeI\ 5198.7 \AA\ and \FeII\ 5197.6 \AA\ lines. The velocities are obtained after $\lambda$-meter technique at the position around the core of each line. The circles mark the distances in the penumbra where we performed  the fit to the velocities.}
\label{fig:contin}
\end{figure*}

\subsection{Sunspots}

Observations were taken at the VTT during the morning of 7th of October, 2012. As a target we used an isolated regular-shaped sunspot belonging to the NOAA 11582 group located at coordinates (914\arcsec,  -216\arcsec) off solar disc center.  By means of the spectrograph of the VTT we scanned the sunspot using several carefully selected spectral intervals.  Each interval was selected from the the list of unblended spectral lines with a clean continuum provided in \cite{Gurtovenko+Kostik1989} under a criterium of containing at least one \FeI\ and one \FeII\ spectral line close in wavelength to fit on the same CCD. We selected pairs of lines with as similar formation heights as possible. The lines are listed in Table 1. The scan step was 0.18\arcsec\ and 250 positions were taken. Different spectral intervals were scanned successively starting from 9:26 UT and it took about 5 min to complete each scan. The seeing conditions were exceptional during the morning, with the $r_0$ parameter varying around 15-20 cm. 

\begin{figure*}[!]
\center
\includegraphics[width=15.0cm]{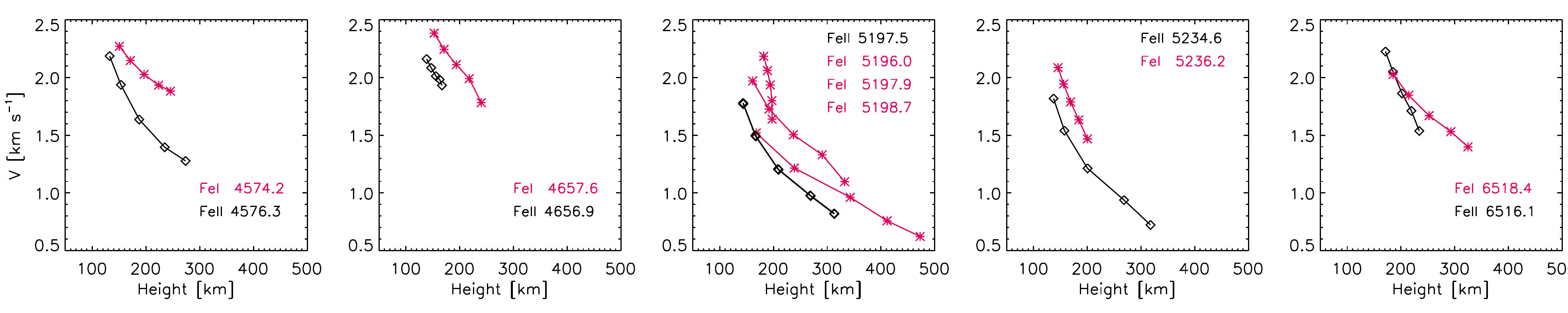}
\caption{$\lambda$-meter velocities as a function of height at the half radius of the visible penumbra, measured at different spectral intervals.}
\label{fig:vlambda}
\end{figure*}

The data were reduced using a standard procedure and velocities were extracted by means of $\lambda$-meter technique \cite{Stebbins+Goode1987}. No absolute wavelength calibration was possible due to the absence of telluric spectral lines in the observed spectral intervals. We performed wavelength calibration by fitting the average quiet Sun spectrum to the FTS atlas.  The 2D images of $\lambda$-meter velocities and continuum intensity were rotated and corrected for the limb perspective. An example of the continuum intensity and velocities of the \FeI\ 5198.7 \AA\ and \FeII\ 5197.6 \AA\ lines are given in Figure \ref{fig:contin}. Our objective was to measure the velocity and the direction of the Evershed flow in the penumbra for neutrals and ions. For that we followed the procedure described in \cite{BellotRubio+etal2003} and fitted the LOS velocity by a function $v_{\rm LOS}(r, \phi)= v_r(r)\sin\theta\cos\phi + v_z(r)\cos\theta$, where $r$ represents the radial distance along penumbra, $\theta$ is the heliocentric angle, $\phi$ is the azimuthal angle around the spot center, and $v_r$ and $v_z$ are the vertical and radial components of the velocity in the local reference frame. The fit was performed for the velocities at six radial distances in the penumbra located at various ranges of $r$, see Figure \ref{fig:contin}. 

The $\lambda$-meter technique allows to obtain the velocity at several positions along the line profile that correspond to different heights in the solar atmosphere. In order to compare \FeI\ and \FeII\ velocities, we need these heights to be reliably calculated. The formation height of spectral lines  can strongly vary at the locations of bright and dark penumbra filaments. We calculated the formation heights of the observed spectral lines in two semi-empirical models of sunspot penumbra, corresponding to a dark filament (model C) and a bright filament (model D) as provided in \cite{Socas-Navarro2007} at the observed heliocentric angle of $\theta=77$\degree\ (Shchukina, private communication).  The height where the optical depth is equal to one at a given wavelength along the line profile was taken to be its formation height. Each $\lambda$-meter velocity was assigned a height calculated this way in the models C and D and then averaged for a given $r$ and $\lambda$-meter position. For a reference, Table 1 provides heights corresponding to the line minimum.  

Figure \ref{fig:vlambda} shows the amplitude of the Evershed flow velocity $v=\sqrt(v_r^2 + v_z^2)$ as a function of height in the model D for all five spectral intervals separately for the \FeI\ lines (red) and \FeII\ lines (black). The results are qualitatively similar for the heights from model C with a different temperature structure. In all the cases the outward velocity decreases with height according to the well known behavior of the Evershed flow. The \FeI\ lines show systematically larger velocities than \FeII\ ones at all spectral intervals. Figure  \ref{fig:vlambda} shows the velocities at the half radius of the visible penumbra. The difference between the \FeI\ and \FeII\ velocities is maximum at radial distances close to the umbra-penumbra boundary and becomes almost negligible at distances close to the visible end of the penumbra. While the results of our analysis are subject to uncertainties due to the height of formation of spectral lines, it is intriguing that all spectral intervals independently show similar behavior with slightly larger velocity of neutral iron at the same photospheric heights. More extensive analysis of these data will be presented in our future publications.

\section{Concluding remarks}

This paper summarizes our recent effort in the theoretical and observational investigation of the effects of partial ionization of the solar plasma into its dynamics and energy balance. We have developed a self-consistent mathematical formulation for the two-fluid and single-fluid description of the partially ionized multi-species solar plasma and its coupling to the radiation field. The single-fluid equations are solved by the extended version of the 3D non-ideal MHD code Mancha. First results of our simulations show the high importance of the ion-neutral effects (ambipolar diffusion) for the heating of the magnetized solar chromosphere. The presence of neutrals is also shown to modify the stability criterium of the Rayleigh-Taylor instability at the border of solar prominences. Observational detection of ion-neutral effects has been attempted. Our preliminary analysis indicates possible differences in the velocities of ions and neutrals in solar prominences and in sunspot penumbra.

\section*{Acknowledgments}   
This work is partially supported by the Spanish Ministry of Science through projects AYA2010-18029 and AYA2011-24808 and by the Leverhulme Trust through project IN-2014-016. This work contributes to the deliverables identified in FP7 European Research Council grant agreement 277829, ``Magnetic connectivity through the Solar Partially Ionized Atmosphere''. The collaboration and discussions with M. Collados, A. D{\'{\i}}az, A. de Vicente, N. Vitas, M. Luna and N. Shchukina are gratefully acknowledged. 

\providecommand{\noopsort}[1]{}\providecommand{\singleletter}[1]{#1}%


%
\end{document}